\documentclass[a4paper,11pt]{article}

\def\Journal#1#2#3#4{{#1} {\bf #2}, #3 (#4)}

\def\NIMA{{\em Nucl. Instrum. Methods} A}

\def\PRL{\em Phys. Rev. Lett.}
\def\PRD{{\em Phys. Rev.} D}

\usepackage{pos}
\usepackage{multirow}
\title{Recent neutrino oscillation result with the IceCube experiment}

\ShortTitle{IceCube 9.3 year muon neutrino disappearance result}

\author{The IceCube Collaboration \\{\normalsize \normalfont(a complete list of authors can be found at the end of the proceedings)}\\}

\emailAdd{shiqiyu@icecube.wisc.edu}

\abstract{

The IceCube South Pole Neutrino Observatory is a Cherenkov detector instrumented in a cubic kilometer of ice at the South Pole. IceCube’s primary scientific goal is the detection of TeV neutrino emissions from astrophysical sources. At the lower center of the IceCube array, there is a subdetector called DeepCore, which has a denser configuration that makes it possible to lower the energy threshold of IceCube and observe GeV-scale neutrinos, opening the window to atmospheric neutrino oscillations studies. Advances in physics sensitivity have recently been achieved by employing Convolutional Neural Networks to reconstruct neutrino interactions in the DeepCore detector. In this contribution, the recent IceCube result from the atmospheric muon neutrino disappearance analysis using the CNN-reconstructed neutrino sample is presented and compared to the existing worldwide measurements.

\vspace{4mm}
{\bfseries Corresponding authors:}
Shiqi Yu$^{1*}$, Jessie Micallef$^{2,3}$\\
{$^{1}$ \itshape Michigan State University}\\
{$^{2}$ \itshape Massachusetts Institute of Technology}\\
{$^{3}$ \itshape Tufts University}\\[4mm]

$^*$ Presenter

\ConferenceLogo{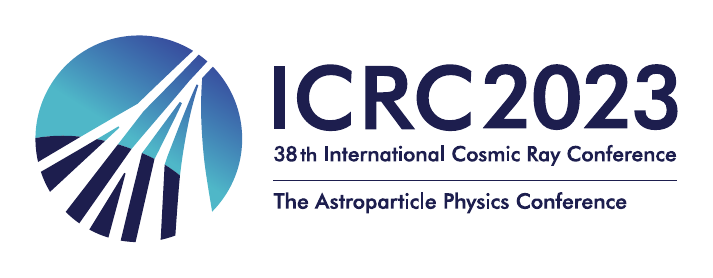}

\FullConference{The 38th International Cosmic Ray Conference (ICRC2023)\\ 26 July -- 3 August, 2023\\ Nagoya, Japan}
}

\begin{document}

\maketitle

\section{Introduction}\label{sec:intro}
Neutrinos are generated and detected as three flavors, namely $\nu_{e,\mu,\tau}$ via weak interactions, while they propagate in their mass eigenstates: $\nu_{1,2,3}$. However, as they travel, they propagate in their respective mass eigenstates, denoted as $\nu_{1,2,3}$. Due to their non-zero masses, the flavor observed in the detection of a neutrino may differ from its original flavor upon creation. This phenomenon is called neutrino oscillation. Neutrinos oscillations have been observed and studied by many experiments~\cite{icecube_prd,icecube_prl,nova_paper,t2k_paper,minos_paper,dayabay_paper}. The probability of being created in one flavor and subsequently detected in another is described by a unitary matrix, the Pontecorvo-Maki-Nakagawa-Sakata (PMNS)~\cite{pmns} matrix. This unitary matrix can be parameterized by three mixing angles ($\theta_{12}$, $\theta_{13}$, and $\theta_{23}$), one CP-violating phase $\delta_{CP}$, and the squared mass differences, $\Delta m_{ij} \equiv m^2_{i}$-$m^2_j$, between the three neutrino masses $m_i$, where $i,j={1,2,3}$. In this study, we measure the values of $\Delta m^2_{32}$ and $\theta_{23}$ via the $\nu_\mu$ disappearance channel by studying atmospheric muon neutrinos with the IceCube Neutrino Observatory. For the other oscillation parameters that are previously well measured, we employ the reported values by~\cite{pdg} ($\Delta m^2_{21} \approx 7.4\times 10^{-5}\rm{eV}^2/c^4$, $\theta_{13}\approx9^{\circ}$, and $\theta_{12}\approx34^{\circ}$).

The spectrum of cosmic rays at Earth follows a power-law with an isotropic distribution. This generates an atmospheric neutrino flux that also follows a power-law for all zenith angles. In this analysis, we make use of the energy of the neutrino, E, and the distance from their generation point to the detector, L, which can be parametrized as a function of the zenith angle, $\theta_{zenith}$.

\begin{figure}[!bt]\centering
\includegraphics[width=0.6\columnwidth]
{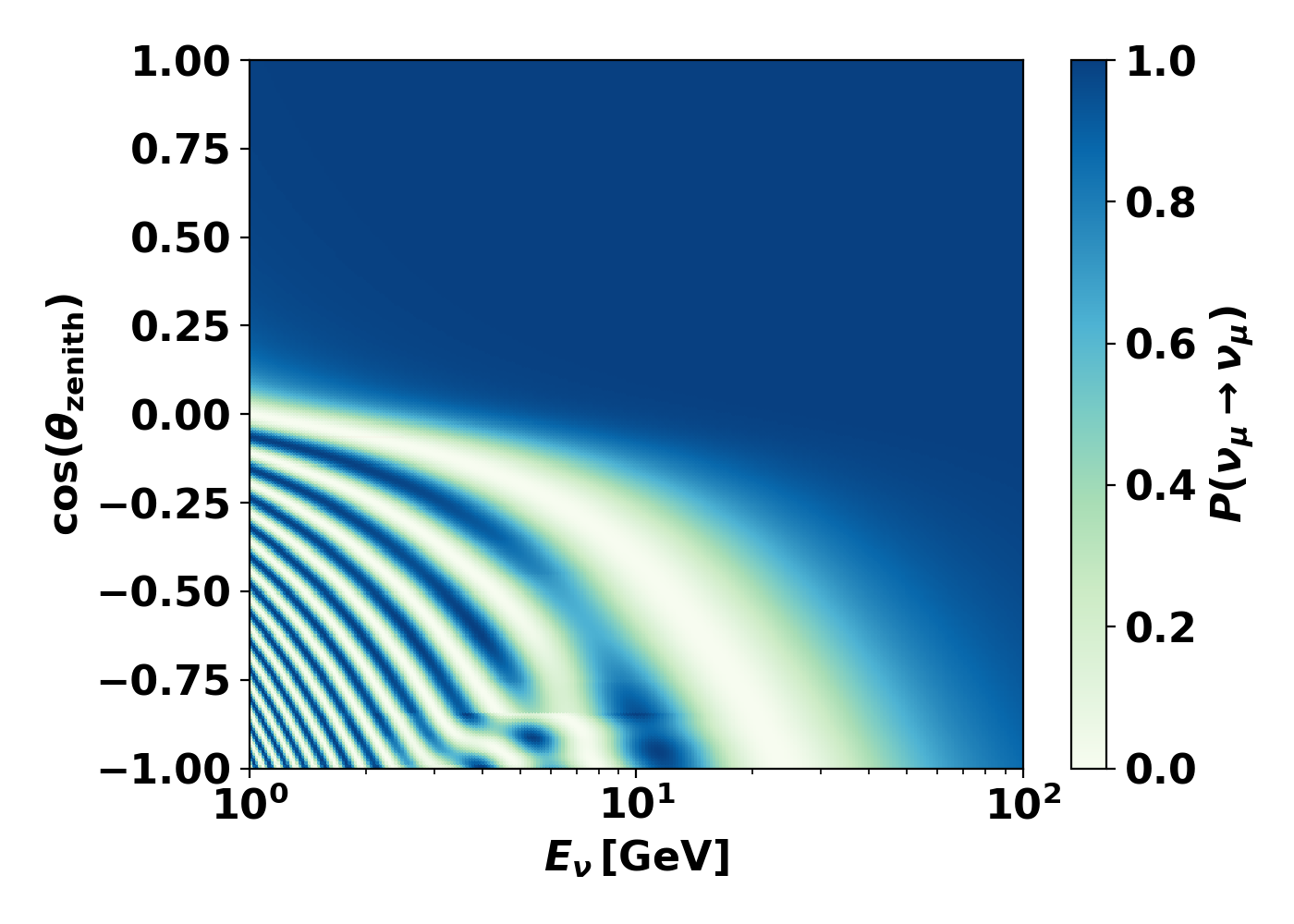}
\caption{\label{fig:osc}Distribution of $\nu_\mu$ survival probability with color representing the value of probability at given values of $\cos(\theta_{\rm{zenith}})$ and E with oscillation parameters from the previous result~\cite{verification_paper}.}
\end{figure}

Atmospheric muon neutrinos are produced by the hadronic processes of cosmic rays interacting with matter in the atmosphere. The interaction spreads all over the atmosphere at all energy ranges, which provides a rich muon neutrino sample with broad ranges of neutrino energy (E) and distance of travel (L). The existing long-baseline experiments have fixed baselines and narrowly peaked neutrino beam energies optimized for neutrino oscillation studies, while IceCube can use the highest L and E in the oscillation analysis. In the effective approximation of two-flavor oscillations, the $\nu_\mu$ survival probability reads: \begin{equation}\label{eq:osc}
P(\nu_\mu \rightarrow \nu_\mu ) \approx 1-\sin^2(2\theta_{23})\sin^2\frac{1.27\Delta m^2_{32}L}{E},
\end{equation}
which depends on the distance traveled, L, and the neutrino energy E. The mixing angle of $\theta_{23}$ and mass splitting of $\Delta m^2_{32}$ are the parameters to be measured and are plotted against L (represented by arrival angle, $\theta_{\rm{zenith}}$) and E in Figure~\ref{fig:osc}. 
The values of $\Delta m^2_{32}$ can affect the frequency of oscillation stripes (see Figure~\ref{fig:osc}), and hence, the locations of the strips. The brightness of the stripes is affected by the value of $\theta_{23}$, which corresponds to the amplitude in Equation~\ref{eq:osc}. The sensitivities to these two oscillation parameters arise mainly from the neutrino sample arriving through the Earth ($\cos\theta_{\rm zenith} \lesssim 0$) with energy between 5 and 100 GeV. Additionally, the first oscillation ``dip'', for example, near 30 GeV in Figure~\ref{fig:osc} with the underlying assumptions on the oscillation parameters, gives us a strong sensitivity to the $\Delta m^2_{32}$ value.

\section{DeepCore Detector}

\begin{figure}[!tb]\centering
\includegraphics[width=0.5\columnwidth, trim=20cm 0cm 0cm 0cm]{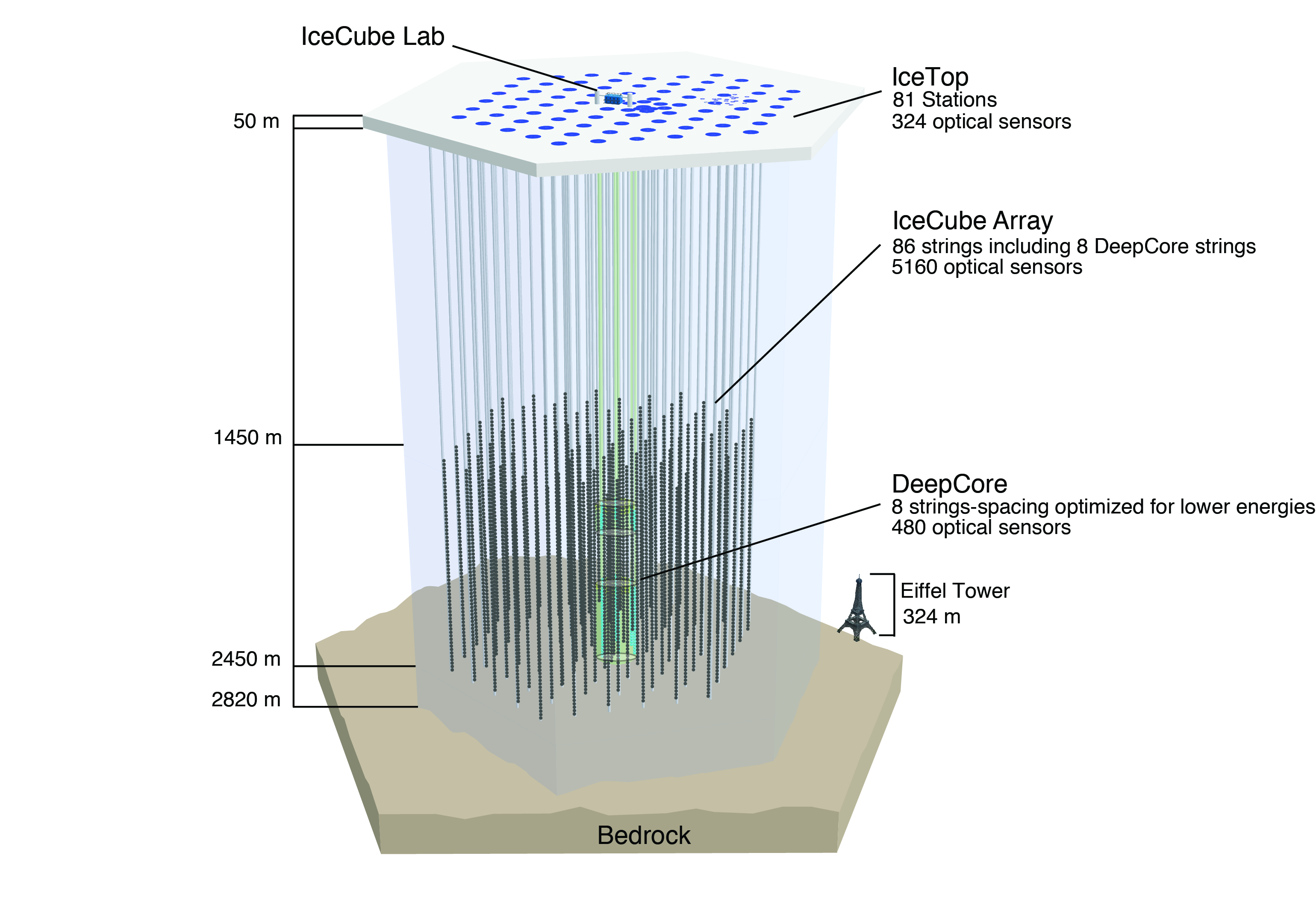}
\includegraphics[width=0.48\columnwidth, trim=0cm 0cm 0cm 40cm]{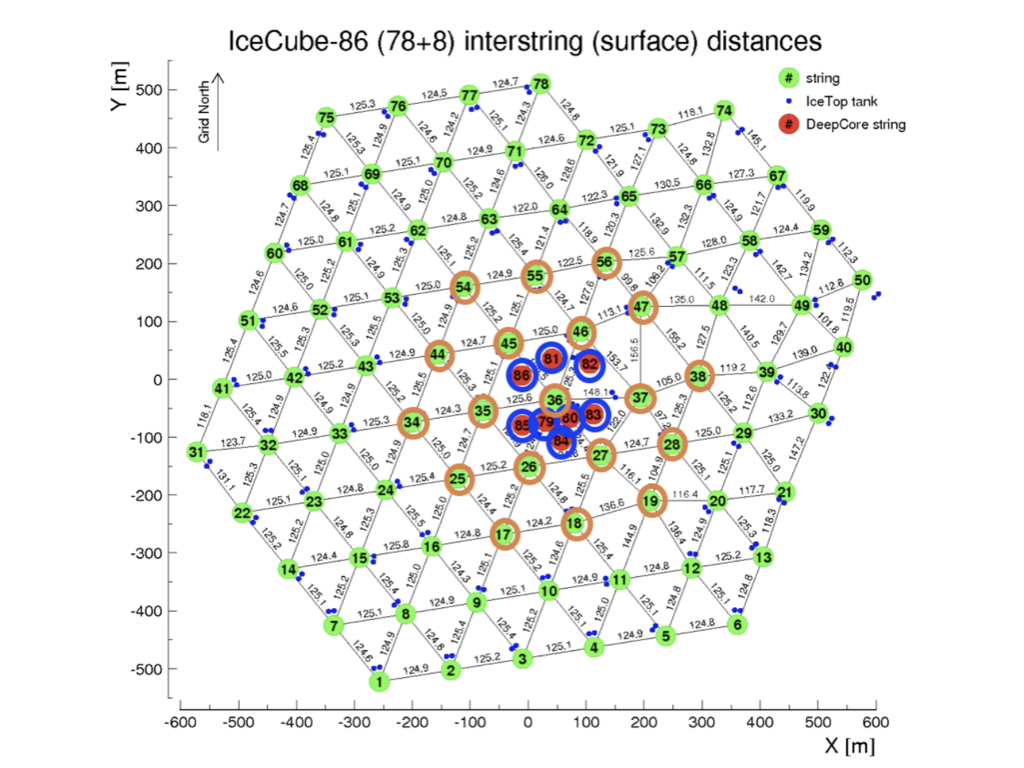}
\caption{\label{fig:detector} IceCube neutrino observatory's main in-ice array and DeepCore sub-detector (left) and surface layout of strings (right) where red-filled DeepCore strings and orange-circled IceCube main strings are used in the CNN reconstruction.}
\end{figure}

The IceCube detector (see Figure~\ref{fig:detector}) comprises 5,160 digital optical modules (DOMs) instrumenting a volume of over one km$^3$ of South Pole glacial ice deep below the surface. Each DOM contains a photomultiplier tube (PMT) and the associated electronics to detect and read out photons as electronic signals. These DOMs can detect Cherenkov photons produced by the relativistic charged particles propagating through the ice, produced initially by neutrino interactions within the detector, and convert these signals into digitized waveforms. The arrival photons' charge and time information can be extracted from the digitized waveforms and used as inputs to a Convolutional Neural Network (CNN) \cite{paper:zenith,paper:energy,paper:nufact} for reconstructing neutrino interactions. The DeepCore sub-detector, located in the bottom center of the IceCube array (see left panel of Figure~\ref{fig:detector}), and the surrounding IceCube strings of the main array (eight red-filled and 19 orange-circled green dots in the right panel of Figure~\ref{fig:detector}) offer exceptional capabilities to reconstruct events in the sub-100 GeV energy range. The DeepCore detector has a denser instrumented volume of ice ($\sim$\.10$^7$ m$^3$) with higher quantum efficiency DOMs compared with those of the main array, which grants us the ability to study neutrinos with observed energy between 5 and 100 GeV arriving through the Earth (with L$\sim1.3\times10^4$ km for those passing through the Earth core). 

This analysis uses the data taken between 2012 and 2021, an equivalent livetime of 9.3 years. The simulation and calibration techniques are the same as the previous result~\cite{verification_paper}, whereas machine-learning (ML) reconstruction techniques are developed and employed, and background-like neutrino candidates help to better constrain systematic uncertainties.


\section{\label{sec:reconstruction} Reconstruction and Event Selection}
We developed and applied the convolutional neural networks (CNNs) focused on the DeepCore sub-detector to reconstruct the sub-100 GeV neutrino sample, which contributes the most to the sensitivity of this study. With the help of CNN reconstructions, we select a final neutrino-rich sample with contamination from atmospheric muons well below 1\% of the selected sample.

\begin{figure*}[tb!]\centering
\includegraphics[width=0.5\columnwidth]
{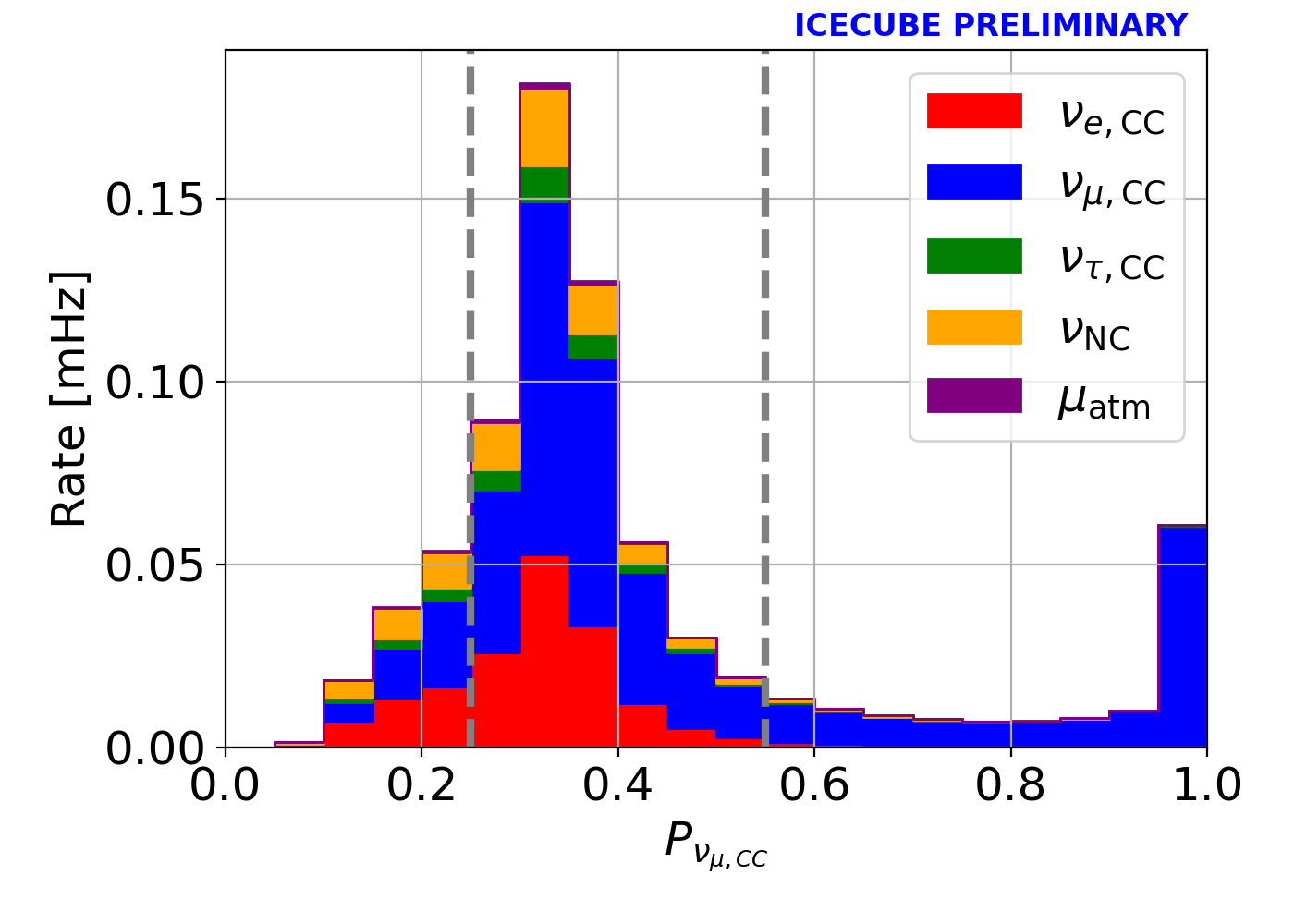}
\caption{\label{fig:pid} Stacked distributions of CNN-reconstructed PID with color representing different Monte Carlo (MC) components, dashed lines indicate boundaries between cascade- (left), mixed- (middle), and signal-like (right) events.}
\end{figure*}

The training of CNNs is optimized separately for neutrino energy~\cite{paper:energy}, arrival direction ($\theta_{\rm{zenith}}$)~\cite{paper:zenith}, interaction vertex position, particle identification (PID), and atmospheric muon classification~\cite{paper:nufact}. We keep the neutrino candidates with their starting vertex close to DeepCore to ensure better reconstruction resolution. Energy and zenith cuts are applied to keep neutrinos with reconstructed energy between 5 and 100 GeV arriving from below the horizon, which is the region, as described in Section~\ref{sec:intro} and shown in Figure~\ref{fig:osc}, that gives us the best sensitivity to oscillation parameter measurements. The CNN-reconstructed PID classifier (as shown in Figure~\ref{fig:pid}) selects the signal-like candidates, i.e., $\nu_\mu$ charged current (CC) interactions, over the background-like neutrino interactions (all remaining types) of this analysis. Signal-like events have a track-like topology in the detector because of their outgoing primary muons, while background-like events look like scattered cascades due to their electromagnetic and hadronic showers. 
A boosted decision tree is employed to serve as the atmospheric muon classifier, which helps to keep the final sample neutrino-dominated with the rate of atmospheric muon background events well within 0.6\% of the sample. 

The CNNs are trained independently using differently optimized samples to achieve optimized reconstruction performances on all variables. The neutrino energy and interaction vertex position are trained on a $\nu_\mu$ CC sample with a uniform energy spectrum between 1 and 300 GeV, with a tail extending to 500 GeV. The zenith angle, $\theta_{\rm{zenith}}$, is trained using $\nu_\mu$ CC events starting and ending near DeepCore generated with a uniform true $\theta_{\rm{zenith}}$ distribution. The PID identifier is trained on a sample with an equal number of track-like and cascade-like neutrino events. The atmospheric muon classifier is trained on a balanced sample of track-like and cascade-like neutrino interactions and atmospheric muons. All the DOMs on the 8 DeepCore and surrounding 19 IceCube strings, as shown in Figure~\ref{fig:detector}, are incorporated to the CNN via two separate input layers due to their different spatial densities. With a similar performance to the state-of-the-art likelihood-based reconstructions~\cite{paper_retro}, the most considerable improvement of using CNN is from the processing speed (approximately 3,000 times faster), which is a significant advantage considering the large statistics of the full MC production of atmospheric neutrino datasets used in these analyses. 

\section{\label{sec:analysis}Analysis}
\begin{figure*}[tb!]
\includegraphics[width=\columnwidth]{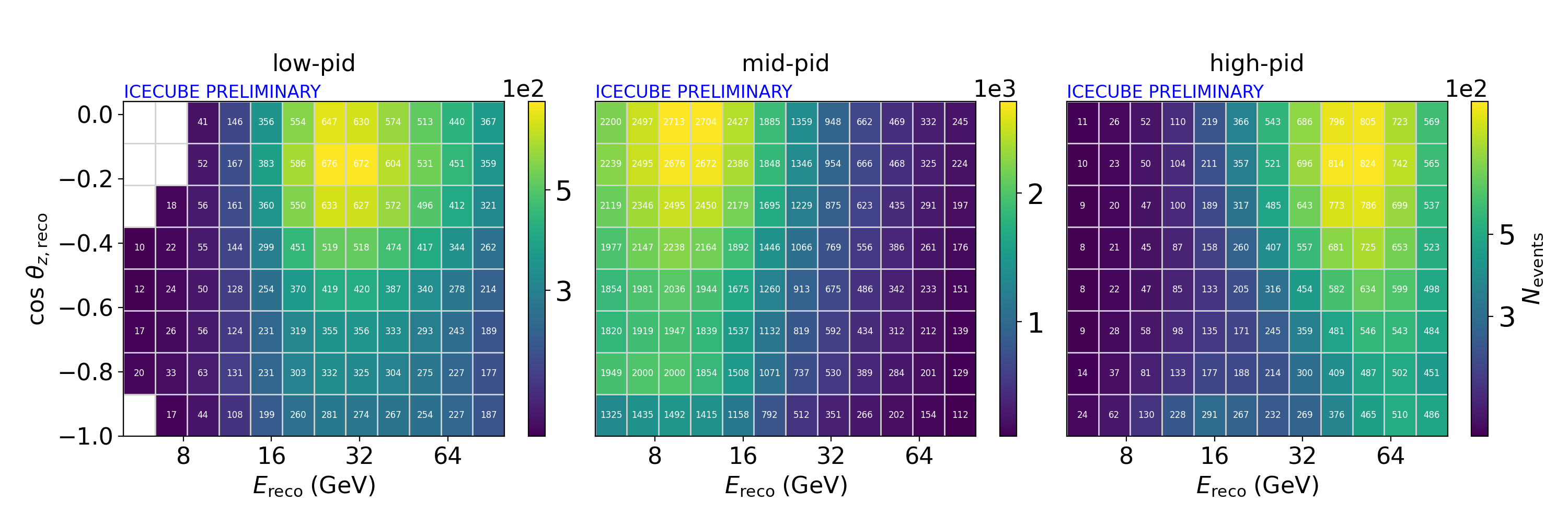}
\caption{\label{fig:nominal} Selected analysis sample in bins of neutrino energy, zenith angle, and PID (cascade-, mixed-, and track-like samples from left to right). Blank bins are not used in the analysis due to their low MC statistics.}
\end{figure*}

We bin the selected neutrino sample using 3D binning: reconstructed energy, $\cos(\theta_{\rm{zenith}})$, and PID (see Figure~\ref{fig:pid}). The high-PID bin (PID $\geq0.55$) has the highest purity of signal-like events, while cascade-like events dominate the low-PID bin (PID $<0.25$). The ten logarithmic energy bins and eight linear $\cos(\theta_{\rm{zenith}})$ bins help to reveal the oscillation pattern in the low energy up-going region while not pushing beyond the limitation of reconstruction resolution. The binned analysis sample can be found in Figure~\ref{fig:nominal}. 

\begin{figure*}[tb!]
\centering\includegraphics[width=0.6\columnwidth]{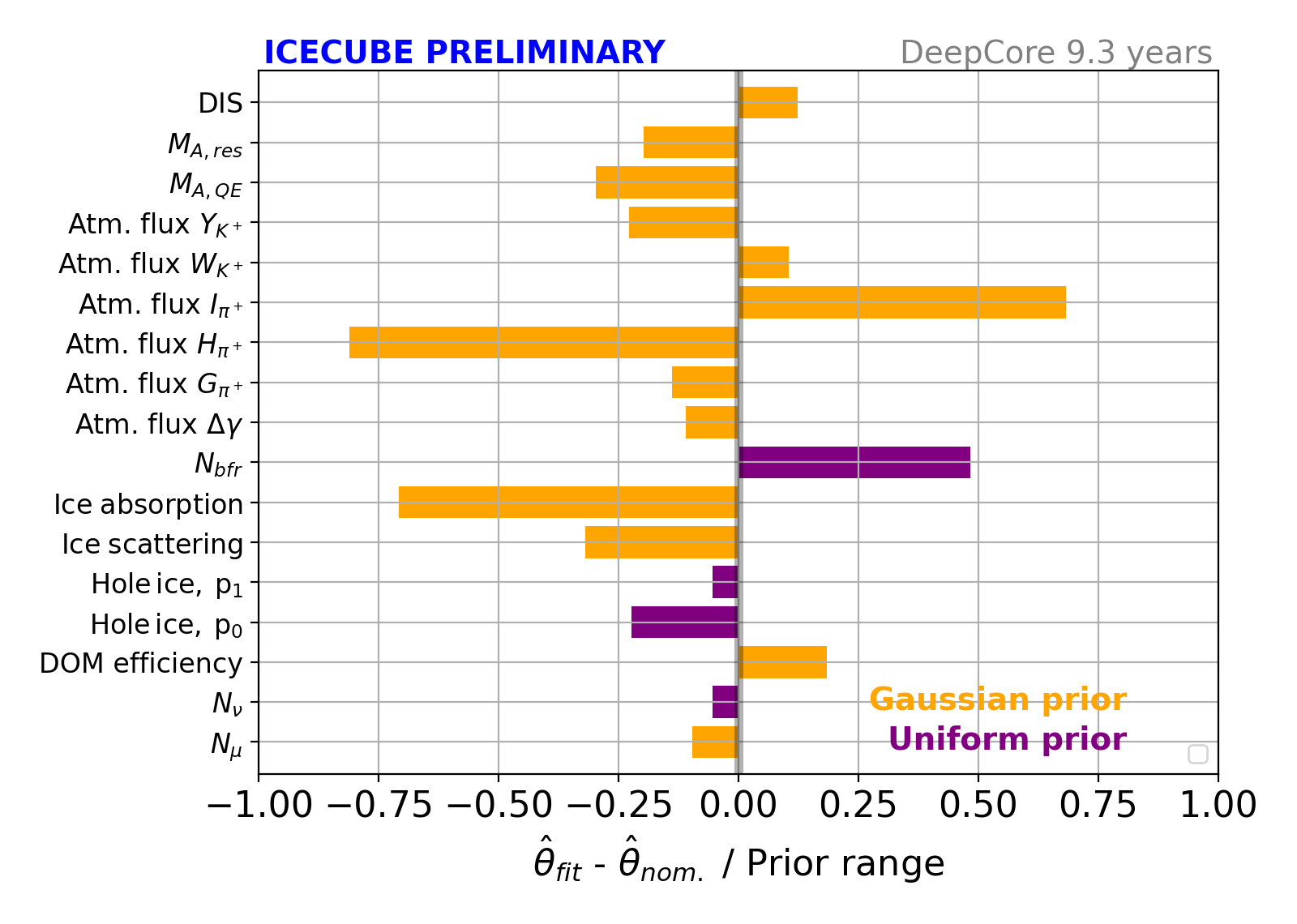}
\caption{\label{fig:sys}Fitted systematic uncertainty parameters pulled from nominal values compared to ranges of priors. Detailed descriptions and references of individual parameters are in the main text.} 
\end{figure*}

The treatment of systemtatic uncertainties follow a similar procedure of the previous analysis.~\cite{verification_paper}. In this analysis, the list of free parameters is decided by re-evaluating their impacts on the recovery of the true underlying physics parameters. The fitted values of the systematic parameters compared to their nominal values and prior ranges and shown in Figure~\ref{fig:sys}.
We employ neutrino interaction cross-section uncertainties from GENIE 2.12.8 ~\cite{paper_genie}, except for the deep-inelastic scattering (DIS) parameter which is interpolated between GENIE and CSMS~\cite{paper_csms} for a higher energy range of coverage; uncertainties of atmospheric flux hadronic production are parameterized following the ``Barr'' parameters (``Atm.~flux Y/W/I/H/G'')~\cite{barr}; ``Atm.~flux $\Delta \gamma$'' represents the uncertainty on the cosmic-ray spectral shape; ``$N_{\rm{bfr}}$'' accounts for the difference in ice properties of employing the birefringent polycrystalline microstructure ice-model~\cite{bfr} and that of the nominal MC (SPICE 3.2.1~\cite{verification_paper}); single-DOM light efficiency, parameters affecting photon propagation in glacial ice (``ice absorption'' and ``ice scattering''), and refrozen ice in drilling holes (``Hole ice, p$_{0(1)}$'') have been introduced in the previous analysis~\cite{verification_paper}; and ``$N_{\nu}$'' (``$N_\mu$'') describes the uncertainty on the normalization of neutrino (muon) flux.

\section{\label{sec:result}Result and Conclusion}
After the final selections, the low-energy atmospheric neutrino dataset taken between 2012-2021 contains 150,257 events. \begin{figure*}[tb!]
\includegraphics[width=0.33\columnwidth]{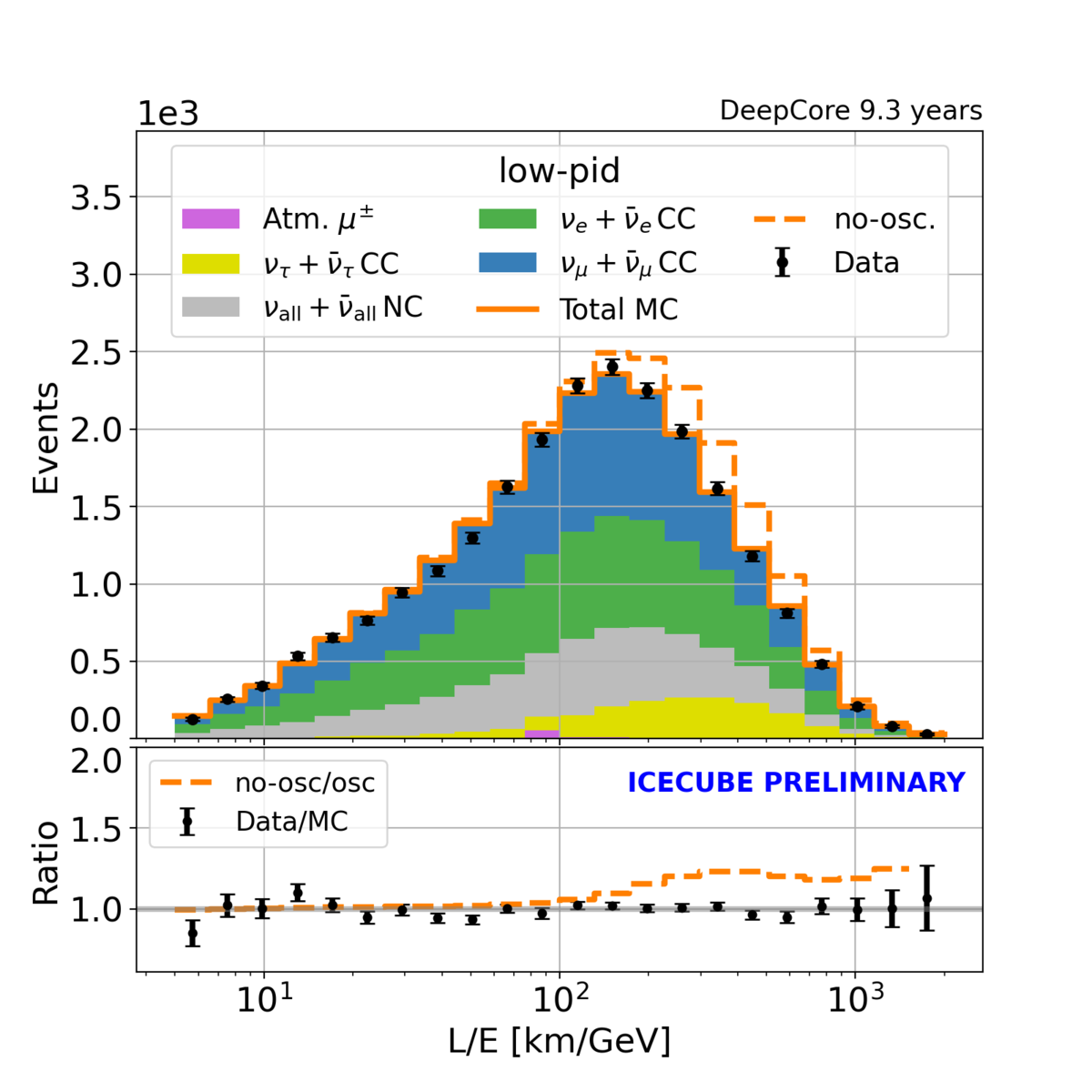}
\includegraphics[width=0.33\columnwidth]{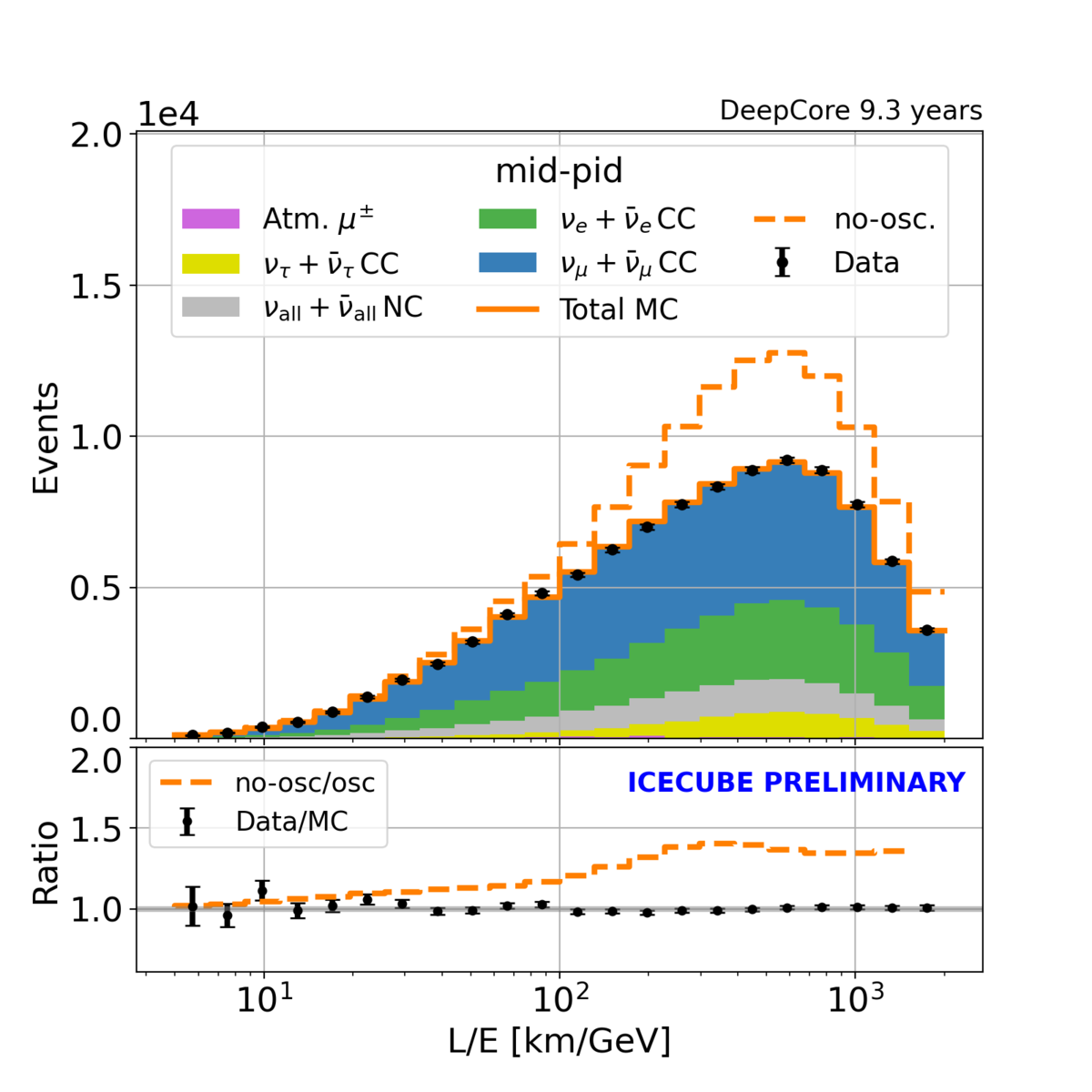}
\includegraphics[width=0.33\columnwidth]{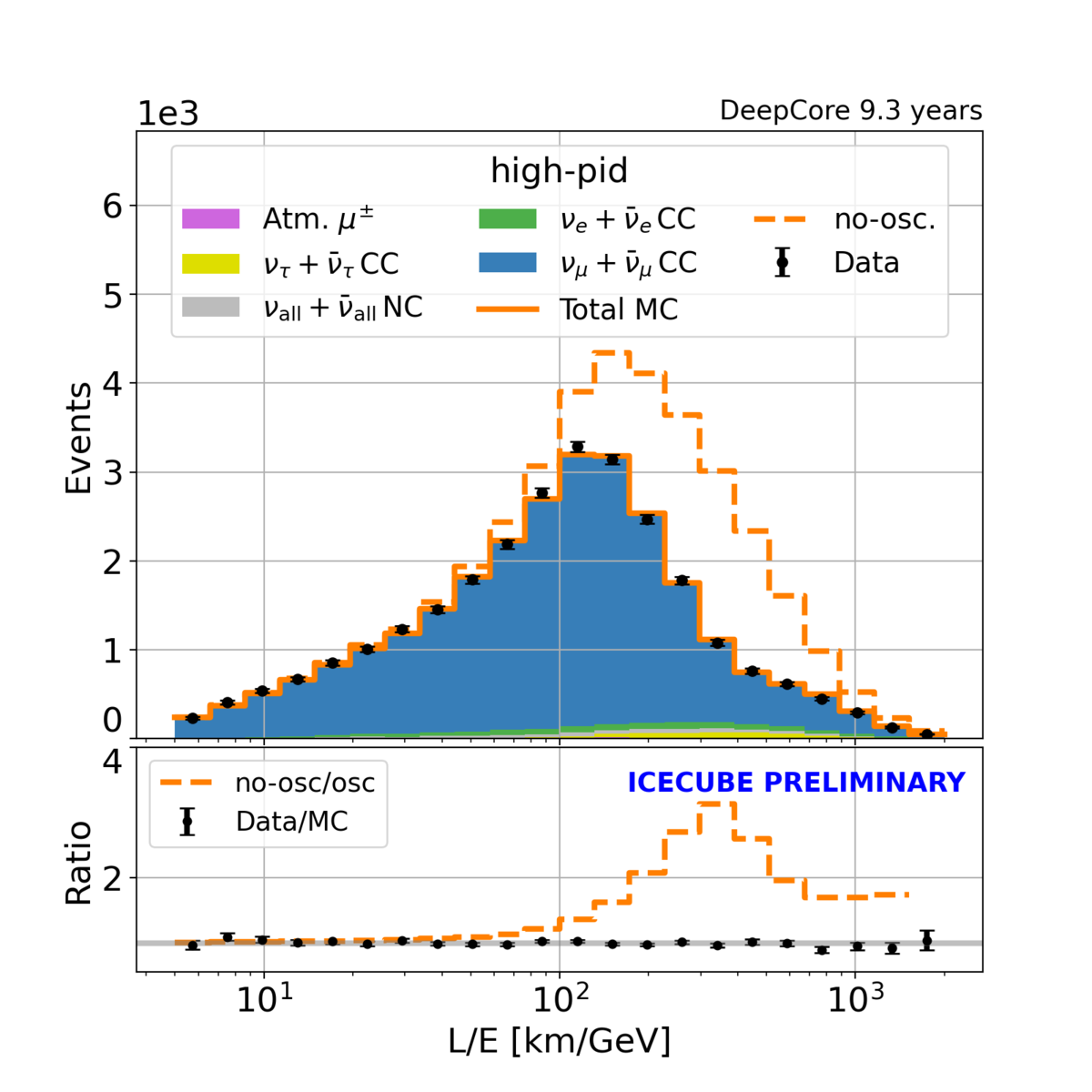}
\caption{\label{fig:lovere}Data (black) and stacked MC comparisons of L/E projections with top panels showing events in cascade- (left), mixed- (middle), and track-like (right) bins with (solid) and without (dashed) muon neutrino disappearance applied, and bottom panels showing ratios of data/MC (black) and MC ratios of without/with oscillations (dashed orange).} 
\end{figure*}
We achieve a good data/MC agreement and a distinctive signature of muon neutrino disappearance in the track-like bin, as shown in Figure~\ref{fig:lovere}.

\begin{figure*}[!tb]
\centering\includegraphics[width=0.5\columnwidth]{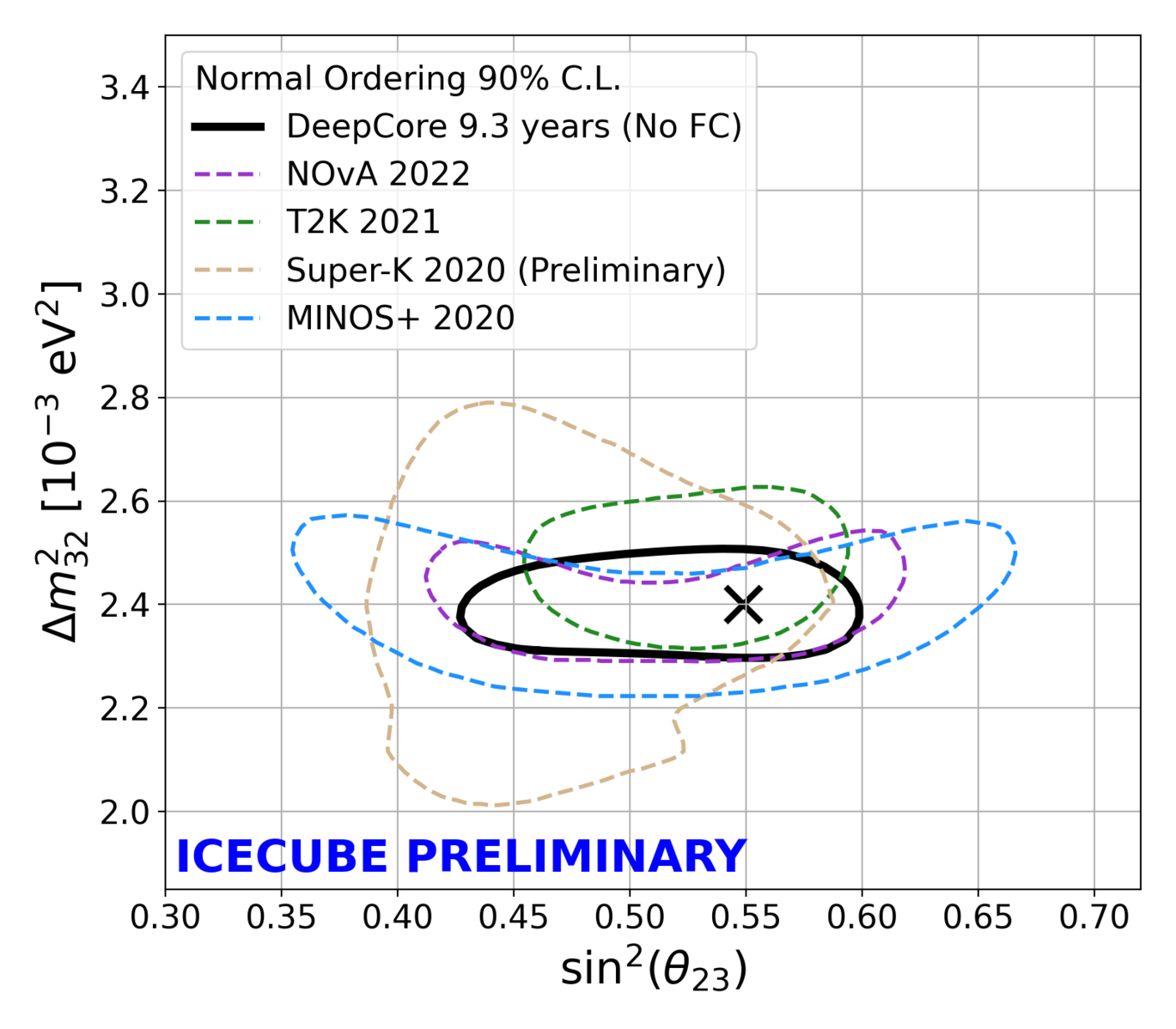}
\caption{\label{fig:contour}90\% C.L. contours (using Wilks' theorem\cite{wilks} and assuming normal mass ordering) and best-fit parameters (cross) of $\sin^2(\theta_{23})$ and $\Delta m^2_{32}$ compared to contours of other experiments~\cite{nova_paper,t2k_paper,minos_paper,superk_paper}.}
\end{figure*}
Figure~\ref{fig:contour} shows the 90\% confidence level (C.L.) contours of $\sin^2(\theta_{23})$ and $\Delta m^2_{32}$ assuming neutrino masses are in normal ordering ($m_3 > m_2 > m_1$). This result is consistent with all the previous accelerator and atmospheric neutrino oscillation measurements, as shown by the strong overlap in the 90\% C.L. contours. Given the competitive sensitivity to the current world-leading measurements, improvements in the precision of global fits to neutrino oscillation parameters are expected once this result is incorporated. Since this analysis is sensitive to a higher energy range compared to other oscillations experiments and the detector technology is unique, it carries a distinct set of systematic uncertainties. The observed consistency is thus a strong validation of the standard three massive neutrino oscillation model.

\begin{figure*}[!tb]
\centering\includegraphics[width=0.45\columnwidth]{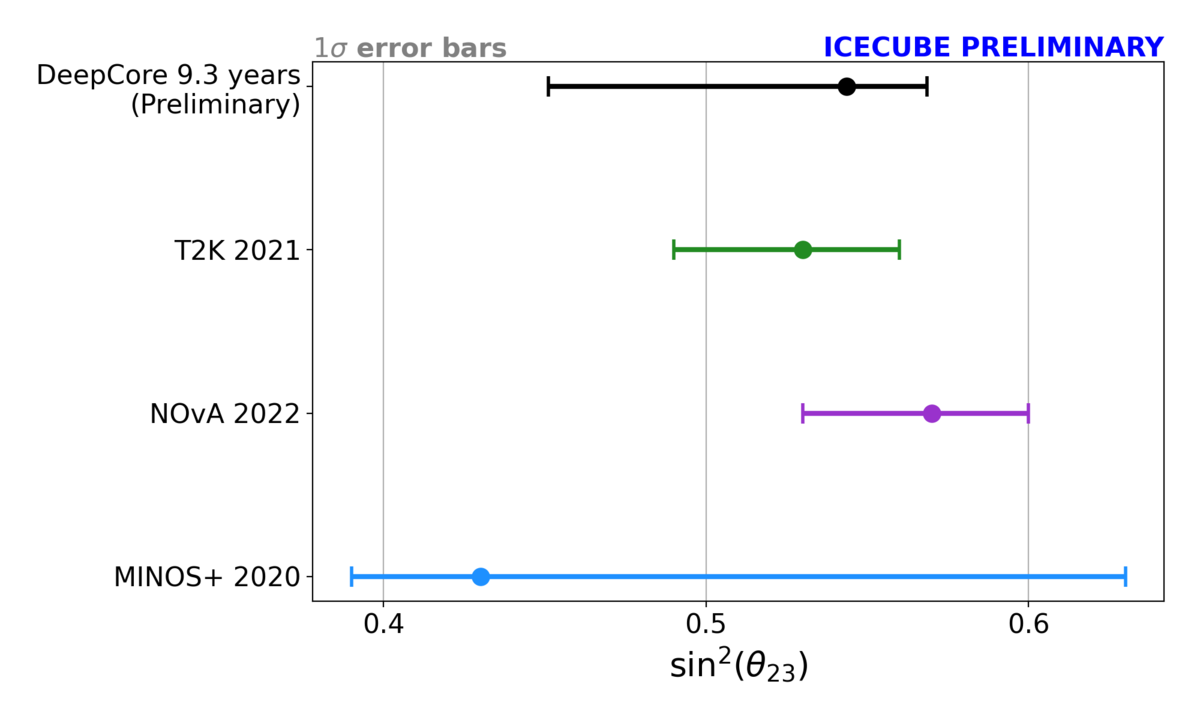}
\includegraphics[width=0.45\columnwidth]{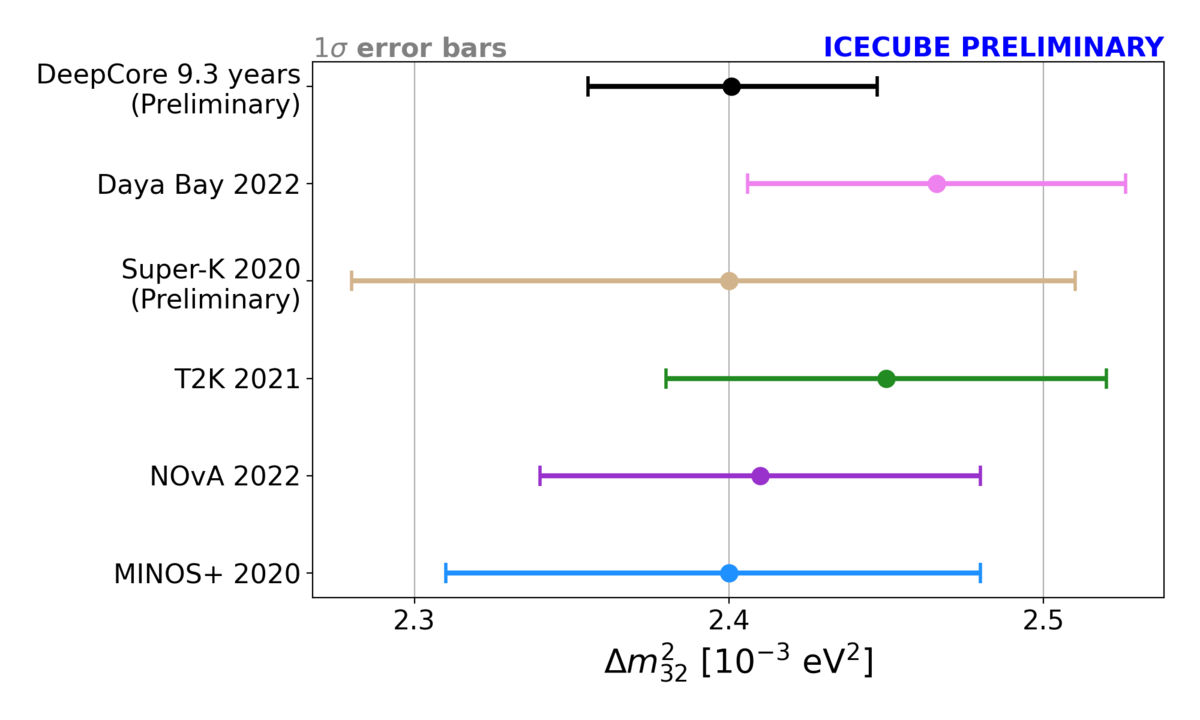}
\caption{\label{fig:errors}One $\sigma$ uncertainties (using Wilks' theorem and assuming normal mass ordering) on $\sin^2(\theta_{23})$ (left) and $\Delta m^2_{32}$ (right) of this result (black) compared with the existing measurements~\cite{nova_paper,t2k_paper,dayabay_paper,minos_paper,superk_talk}.} 
\end{figure*}
As shown in Figure~\ref{fig:errors}, the reported 1$\sigma$ uncertainties on the values of $\sin^2(\theta_{23})$ and $\Delta m^2_{32}$ from different experiments are largely well agreed. The uncertainty of $\Delta m^2_{32}$ measurement from this analysis has a narrower uncertainty than the existing measurements. This is primarily benefited from the first oscillation ``dip'' in our 2D oscillation pattern, as discussed in Section~\ref{sec:intro} and shown in Figure~\ref{fig:osc}, the location of which contributes most to the sensitivity of $\Delta m^2_{32}$ constraint. Meanwhile, it also benefited from the large-statistic of selected neutrino-rich final sample, improved detector calibration and MC models, and the new ML-based reconstruction techniques.

There is a lot of room for future improvements in muon neutrino disappearance measurement using low-energy atmospheric data with IceCube DeepCore. The near-future IceCube Upgrade~\cite{upgrade} will help to further improve our sensitivity to muon neutrino disappearance by improving detector calibration and event resolution. There are improved MC models underway, which better describe the properties of glacial ice and the composition of atmospheric fluxes and can potentially improve future analysis. There are also ongoing analyses that are benefited from the CNN reconstructions and selections, such as non-standard neutrino interaction measurement and neutrino mass ordering searches. The CNN method could also be adapted and applied to the IceCube Upgrade, further improving the precision of neutrino oscillation measurements.


%

\clearpage

\section*{Full Author List: IceCube Collaboration}

\scriptsize
\noindent
R. Abbasi$^{17}$,
M. Ackermann$^{63}$,
J. Adams$^{18}$,
S. K. Agarwalla$^{40,\: 64}$,
J. A. Aguilar$^{12}$,
M. Ahlers$^{22}$,
J.M. Alameddine$^{23}$,
N. M. Amin$^{44}$,
K. Andeen$^{42}$,
G. Anton$^{26}$
% ...
\\
\\
% ...
$^{12}$ Universit{\'e} Libre de Bruxelles, Science Faculty CP230, B-1050 Brussels, Belgium \\
% ...
$^{17}$ Department of Physics, Loyola University Chicago, Chicago, IL 60660, USA \\
$^{18}$ Dept. of Physics and Astronomy, University of Canterbury, Private Bag 4800, Christchurch, New Zealand \\
% ...
$^{22}$ Niels Bohr Institute, University of Copenhagen, DK-2100 Copenhagen, Denmark \\
$^{23}$ Dept. of Physics, TU Dortmund University, D-44221 Dortmund, Germany \\
% ...
$^{26}$ Erlangen Centre for Astroparticle Physics, Friedrich-Alexander-Universit{\"a}t Erlangen-N{\"u}rnberg, D-91058 Erlangen, Germany \\
% ...
$^{40}$ Dept. of Physics and Wisconsin IceCube Particle Astrophysics Center, University of Wisconsin{\textendash}Madison, Madison, WI 53706, USA \\
% ...
$^{42}$ Department of Physics, Marquette University, Milwaukee, WI, 53201, USA \\
% ...
$^{44}$ Bartol Research Institute and Dept. of Physics and Astronomy, University of Delaware, Newark, DE 19716, USA \\
% ...
$^{63}$ Deutsches Elektronen-Synchrotron DESY, Platanenallee 6, 15738 Zeuthen, Germany  \\
$^{64}$ Institute of Physics, Sachivalaya Marg, Sainik School Post, Bhubaneswar 751005, India \\
% ...

\subsection*{Acknowledgements}

\noindent
The authors gratefully acknowledge the support from the following agencies and institutions:
USA {\textendash} U.S. National Science Foundation-Office of Polar Programs,
U.S. National Science Foundation-Physics Division,
U.S. National Science Foundation-EPSCoR,
Wisconsin Alumni Research Foundation,
Center for High Throughput Computing (CHTC) at the University of Wisconsin{\textendash}Madison,
Open Science Grid (OSG),
Advanced Cyberinfrastructure Coordination Ecosystem: Services {\&} Support (ACCESS),
Frontera computing project at the Texas Advanced Computing Center,
U.S. Department of Energy-National Energy Research Scientific Computing Center,
Particle astrophysics research computing center at the University of Maryland,
Institute for Cyber-Enabled Research at Michigan State University,
and Astroparticle physics computational facility at Marquette University;
Belgium {\textendash} Funds for Scientific Research (FRS-FNRS and FWO),
FWO Odysseus and Big Science programmes,
and Belgian Federal Science Policy Office (Belspo);
Germany {\textendash} Bundesministerium f{\"u}r Bildung und Forschung (BMBF),
Deutsche Forschungsgemeinschaft (DFG),
Helmholtz Alliance for Astroparticle Physics (HAP),
Initiative and Networking Fund of the Helmholtz Association,
Deutsches Elektronen Synchrotron (DESY),
and High Performance Computing cluster of the RWTH Aachen;
Sweden {\textendash} Swedish Research Council,
Swedish Polar Research Secretariat,
Swedish National Infrastructure for Computing (SNIC),
and Knut and Alice Wallenberg Foundation;
European Union {\textendash} EGI Advanced Computing for research;
Australia {\textendash} Australian Research Council;
Canada {\textendash} Natural Sciences and Engineering Research Council of Canada,
Calcul Qu{\'e}bec, Compute Ontario, Canada Foundation for Innovation, WestGrid, and Compute Canada;
Denmark {\textendash} Villum Fonden, Carlsberg Foundation, and European Commission;
New Zealand {\textendash} Marsden Fund;
Japan {\textendash} Japan Society for Promotion of Science (JSPS)
and Institute for Global Prominent Research (IGPR) of Chiba University;
Korea {\textendash} National Research Foundation of Korea (NRF);
Switzerland {\textendash} Swiss National Science Foundation (SNSF);
United Kingdom {\textendash} Department of Physics, University of Oxford.

\end{document}